\documentclass[11pt, twocolumn, a4paper]{article} 

\usepackage{graphicx}

\title{Galileo's speedometer: an approach to the concept of\\ instantaneous velocity}

\begin{document}

\author{C.~E.~Aguiar$^1$, M.~F.~Barroso$^1$, P.~M.~C.~Dias$^1$ and M.~F.~B.~Francisquini$^2$ \\
$^1$Instituto de F\'\i sica, Universidade Federal do Rio de Janeiro, RJ, Brasil \\
$^2$Instituto Federal de Educa\c{c}\~ao, Ci\^encia e Tecnologia do Rio de Janeiro, RJ, Brasil}

\date{}

\maketitle

\begin{abstract}
Difficulties presented by students on the concept of instantaneous velocity are well known. This is in part due to instantaneous speed being often defined in terms of the notion of mathematical limit, which may not be clear to many students in introductory physics courses. In this work we present a complementary teaching proposal that can help students to get a better grasp of instantaneous velocity. The approach is based on Galileo's ideas on instantaneous speed, which he borrowed from the Mertonian scholars much before infinitesimal calculus was developed. 
\end{abstract}

\section{Introduction}
The concept of instantaneous velocity is far from trivial to students in introductory physics courses \cite{Trowbridge, Halloun}. Many find it difficult to conform the idea of velocity at a single instant to the common sense notion that a finite distance and a finite time interval are needed in order to compute or even talk about velocity, failing to differentiate between instantaneous and average velocity.  
The usual presentation of instantaneous velocity---the limit of average velocity as the elapsed time goes to zero---may not help alleviating this conceptual obstacle, particularly for students of non-calculus courses. Being unfamiliar with the meaning of such limit, the students are often left with the impression that instantaneous velocity is just a special case of average velocity, good for tiny but still finite time intervals. 

\begin{figure}[htb]
\centering
\includegraphics[width=0.8\linewidth]{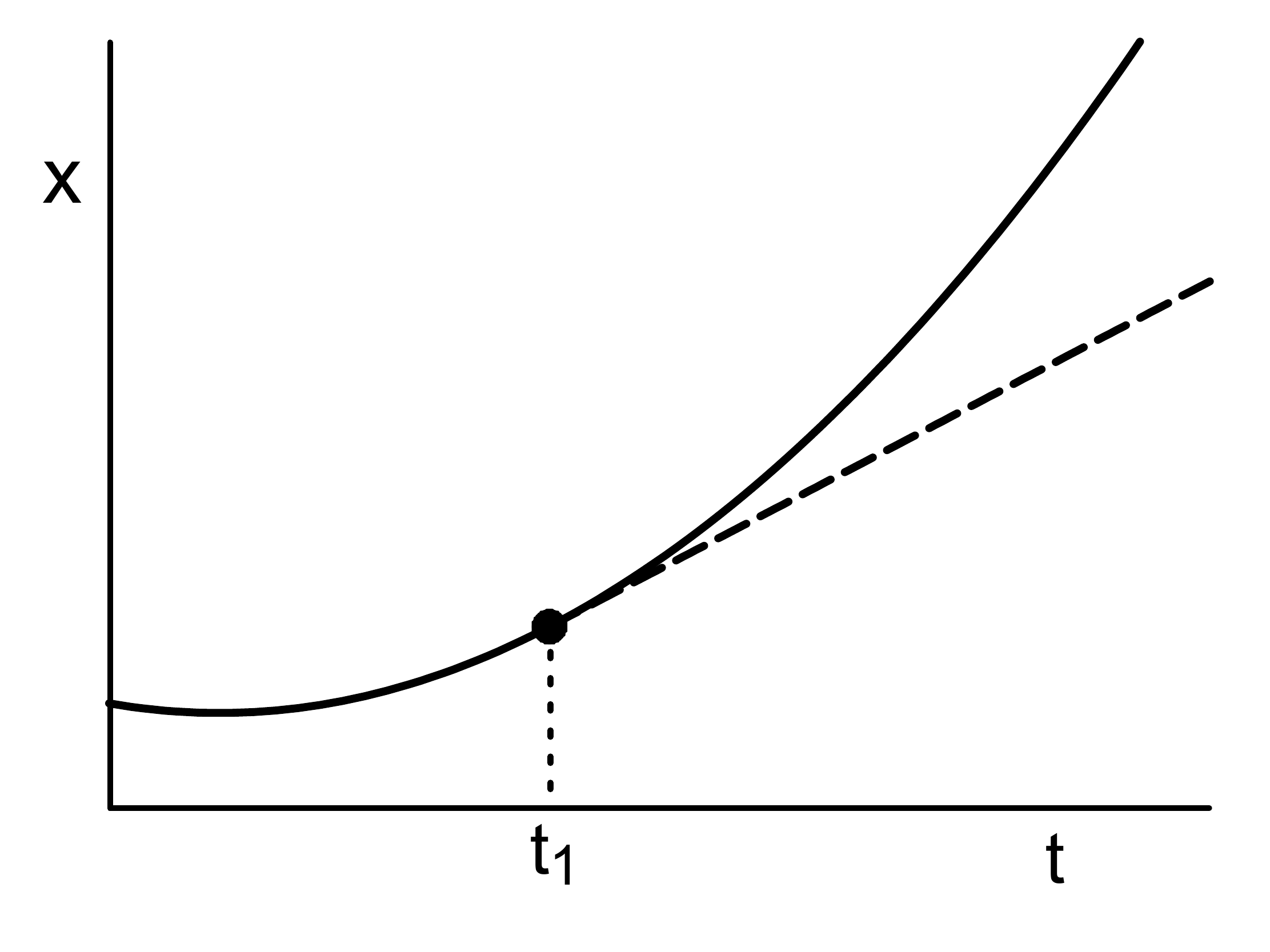}
\caption{The Mertonian definition of instantaneous speed. The solid line represents the position $x$ of a particle in non-uniform motion, as a function of time $t$. The long-dashed line is the path that would be described by the particle if its motion became uniform at instant $t_1$.}
\label{merton-v}
\end{figure}

The idea that velocity refers only to motion across an extended interval goes back to Aristotle. 
The Aristotelian concept of velocity was based on a definition of the `quicker' of two bodies: this would be the one that traverses more space in the same time or, alternatively, that traverses the same space in less time \cite[pp~15--17]{Damerow}. 
In that framework, velocity is a global measure of motion over a finite extension of space and time, akin to the modern average velocity. 
Only in the 14th century was the concept of instantaneous velocity given a clear formulation, which allowed non-uniform motion to be analysed in ways that were ``exact and substantially correct'' \cite[p~20]{Moody}. These advancements were initially carried out by a group of scholars at Merton College, Oxford, collectively known as the Mertonians or Oxford `calculators' \cite{Clagett}. 
The Mertonians characterized instantaneous velocity in the following way: if a body is in non-uniform motion, its instantaneous velocity at a given moment is determined by the path this body \emph{would} describe if it were moved \emph{uniformly} for some time with the same velocity it has at the assigned instant. A graphical depiction of the basic idea is shown in figure~\ref{merton-v}, where the position $x$ of a particle in non-uniform motion is given as function of time $t$ (a plot the Mertonians did not make but is convenient to us). The instantaneous velocity at time $t_1$ was defined as the velocity of the uniform motion which would arise if one imagines that the acceleration were suddenly `switched off' at the moment $t_1$. The instantaneous velocity at $t_1$ is then given by  $\Delta x / \Delta t$ taken over any (finite) span of the uniform motion.
It should be noted that this is a non-calculus definition of instantaneous velocity, formulated long before infinitesimal calculus was developed. 

\begin{figure}[htb]
\centering
\includegraphics[width=0.95\linewidth]{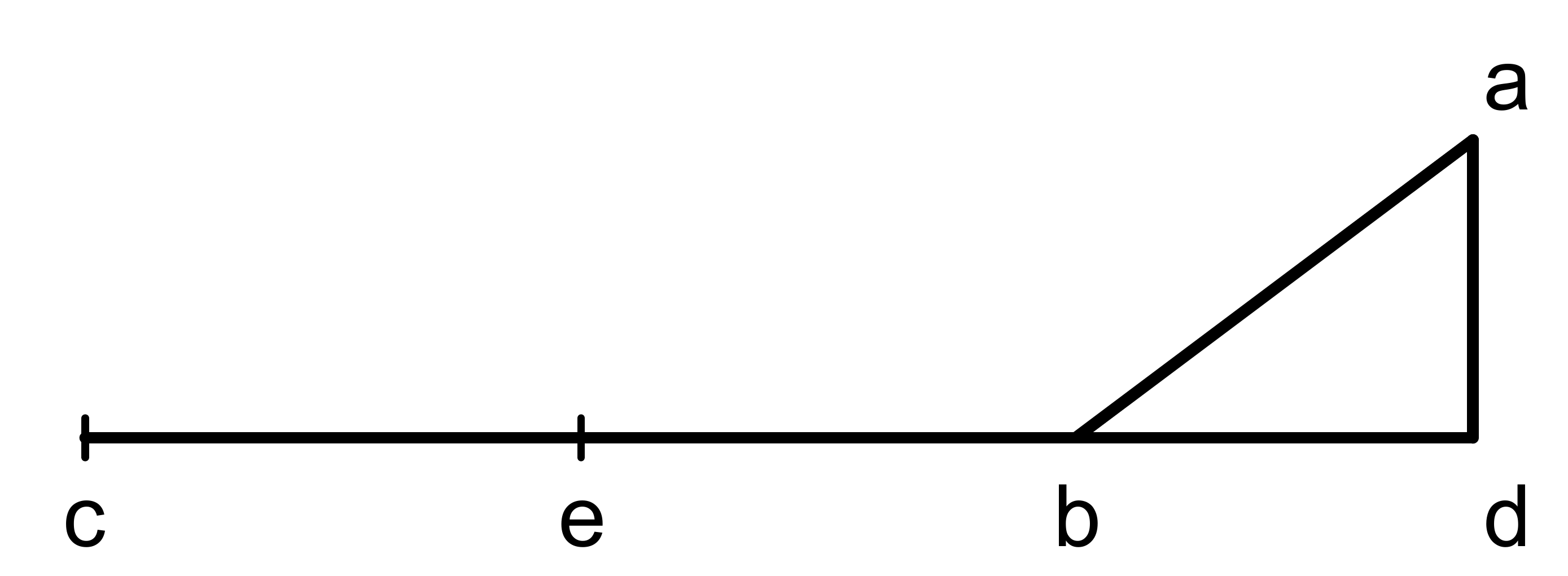}
\caption{Inclined plane with horizontal extension, as described by Galileo in his folio~163v, circa 1602.}
\label{incline}
\end{figure}

In the early 1600s Galileo applied the Mertonian definition of instantaneous speed to a physically realizable system: an inclined plane with a horizontal extension. The device is illustrated in figure \ref{incline},  copied from a drawing made by Galileo on his folio~163v \cite[p~361]{Damerow}, dated around 1602. 
The non-uniform motion to be analysed with this gadget is the accelerated descent of a sphere along the incline $ab$. The uniform motion conceived by the Mertonians (the long-dashed line in figure~\ref{merton-v}) is made real by the horizontal sector $bc$ of the track, on which, as  was already known by Galileo, the sphere moves uniformly~\cite[p~128]{Drake}. 

Galileo's device provides a way to measure the instantaneous speed at the end point $b$ of the incline. If $T$ is the time it takes for the sphere to cross a distance $D$ on the horizontal track, one obtains from the Mertonian definition that the instantaneous speed at the end of the ramp is $D/T$. In a sense, knowledge of the inertia principle showed Galileo how he could switch off the acceleration of a non-uniform motion and allowed him to convert the Mertonian abstract concept of instantaneous velocity into an operational definition. For this reason we have dubbed the device of figure~\ref{incline} ``Galileo's speedometer''. 

In this paper we suggest that Galileo's speedometer can be a useful tool for teaching the concept of instantaneous speed. Presented together with the Mertonian definition, it may be a helpful addition to the customary introduction of instantaneous speed as a limit of the average speed, especially for students not familiar with calculus. 
Emphasising from the beginning that instantaneous speed is assigned to a single instant, not to a finite time interval (and that these should be distinguished, see~\cite[p~25]{Arons}), this framework is less prone to reinforce the misconception that average and instantaneous speeds are basically the same thing \cite{Halloun}. 
Moreover, Galileo's device provides an opportunity to engage students in a practical activity, reifying an abstract concept that many of them find difficult to grasp.
In the next sections we describe in some detail how measurements of instantaneous velocity are carried out with the speedometer and the results that are obtained with it.

\section{Galileo's speedometer and the double-distance rule}

\begin{figure}[htb]
\centering
\includegraphics[width=0.9\linewidth]{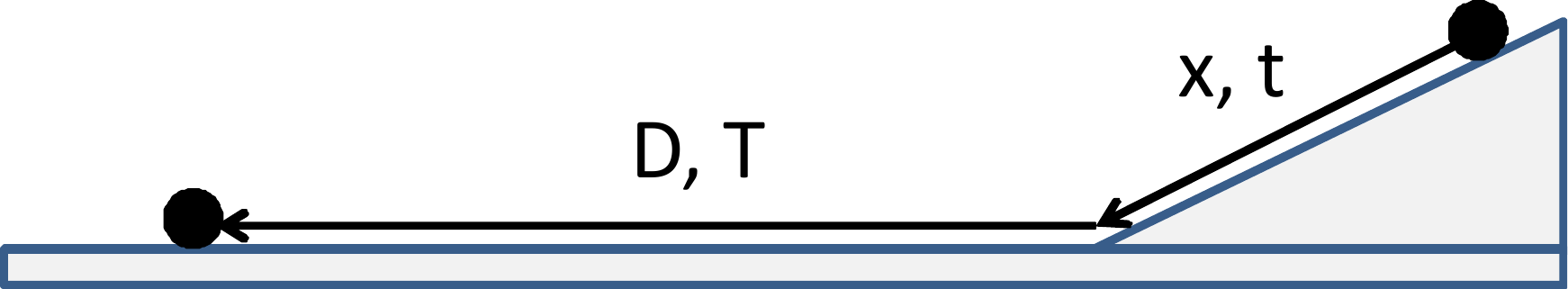}
\caption{Galileo's speedometer. A sphere starting from rest traverses distances $x$ and $D$ on the inclined and horizontal tracks, in time intervals $t$ and $T$.}
\label{aceler}
\end{figure}

Figure \ref{aceler} indicates the relevant quantities in a typical experiment with the speedometer. A sphere starts from rest on the incline and travels in time $t$ a distance $x$ to the bottom of the ramp. It is then deflected into the horizontal track,\footnote{
In order to avoid the bouncing of the sphere at the junction of the inclined and horizontal tracks, we have bent a straight rail so that the transition between the two sectors is a small curved path. } 
where it runs a distance $D$ in time $T$. The values of $x$ and $D$ are chosen at will and the corresponding time intervals $t$ and $T$ are measured. 
Using the Mertonian definition, the instantaneous speed at the foot of the incline is $v = D/T$. 
Repeating the process for different initial points on the incline, the values of $t$, $x$ and $v$ can be assembled in a table.

We performed the time measurements by video analysis with the \emph{Tracker} software \cite{Tracker}. 
The videos were filmed with a cellphone camera at 120~frames/second, allowing for a time resolution of about $10^{-2}$~s.
There are, of course, many other ways of timing the motion on the track, from photogates to recording the sound of little bells hung over the sphere's path. The latter procedure has been applied quite successfully in high-school classes~\cite{Silva}.

\begin{figure}[htb]
\centering
\includegraphics[width=0.8\linewidth]{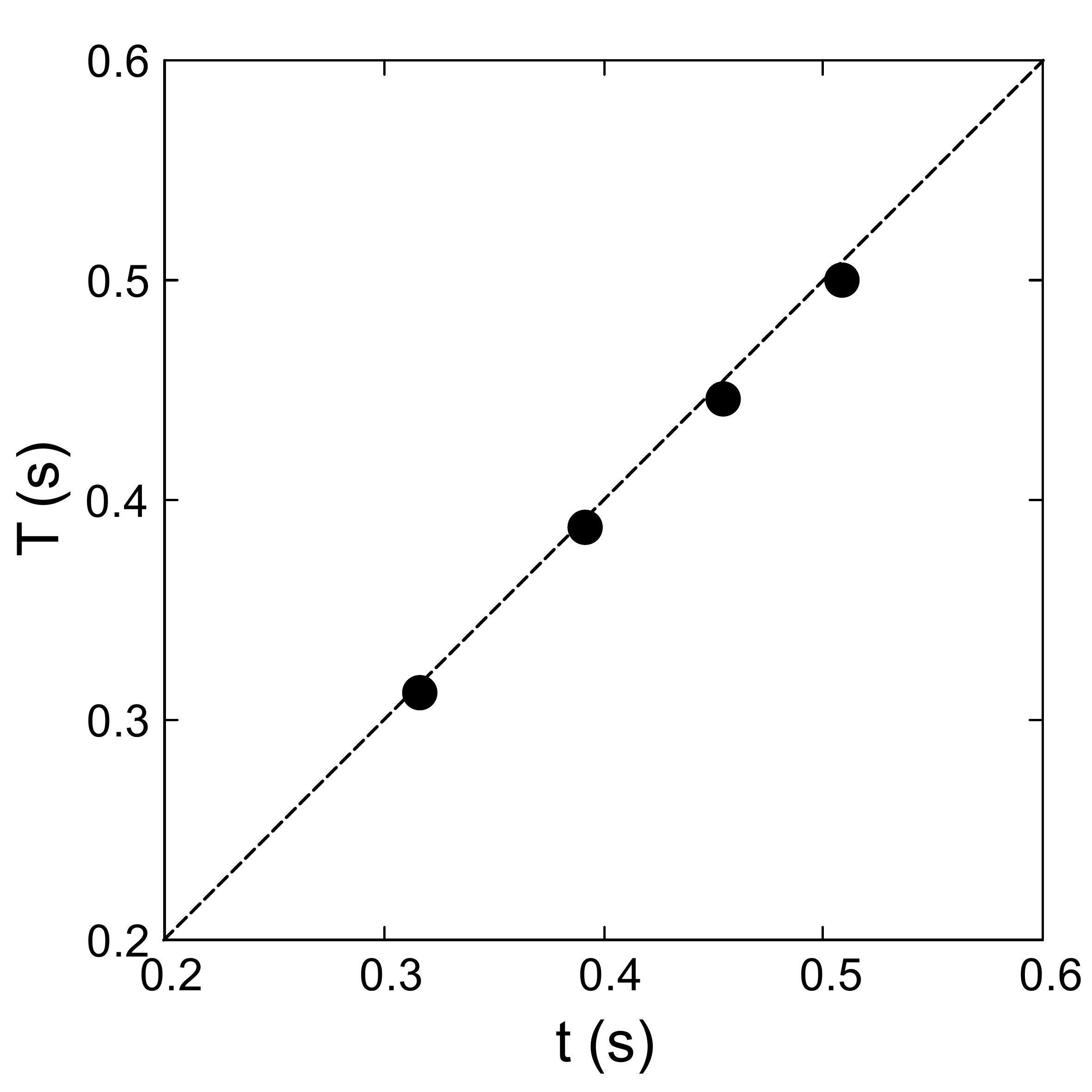}
\caption{Test of the double-distance rule for different starting points on the incline. The dashed line corresponds to equal times $T=t$.}
\label{rdd}
\end{figure}

The number of time measurements is reduced by half if one employs yet another advance provided by the Mertonians and extensively used by Galileo, the `double-distance rule'~\cite[ch~3]{Damerow}.  
The rule states that if $D=2x$ (hence the double distance) then  $T=t$. 
It is valid for bodies uniformly accelerated from rest and is easily demonstrated from basic kinematics, being a corollary to the mean speed theorem. 
If the students using the speedometer do not have yet the necessary requisites to follow a demonstration of the double-distance rule, it may be presented to them as an empirical result, obtained by measuring $T$ and $t$ for distances such that $D=2x$ and checking that they are equal. 
A test of the rule is presented in figure~\ref{rdd}, showing that it works rather well. 
No matter how it is introduced, the double-distance rule has a practical consequence for Galileo's speedometer: if we choose $D$ to be $2x$, only the time interval $T$ on the horizontal needs measuring---the descent time $t$ has the same value. The data shown in next section were taken in this way.

Besides the practical value described above, the double-distance rule is important in itself: it was the only way Galileo had to obtain $x$, $t$, $v$ in uniformly accelerated motion from what he knew, uniform motion. This was one of the `tools for thinking' that ultimately led him to the laws of free fall~\cite{Dias}.

\section{Measurements of instantaneous speed}

\begin{figure}[htb]
\centering
\includegraphics[width=0.9\linewidth]{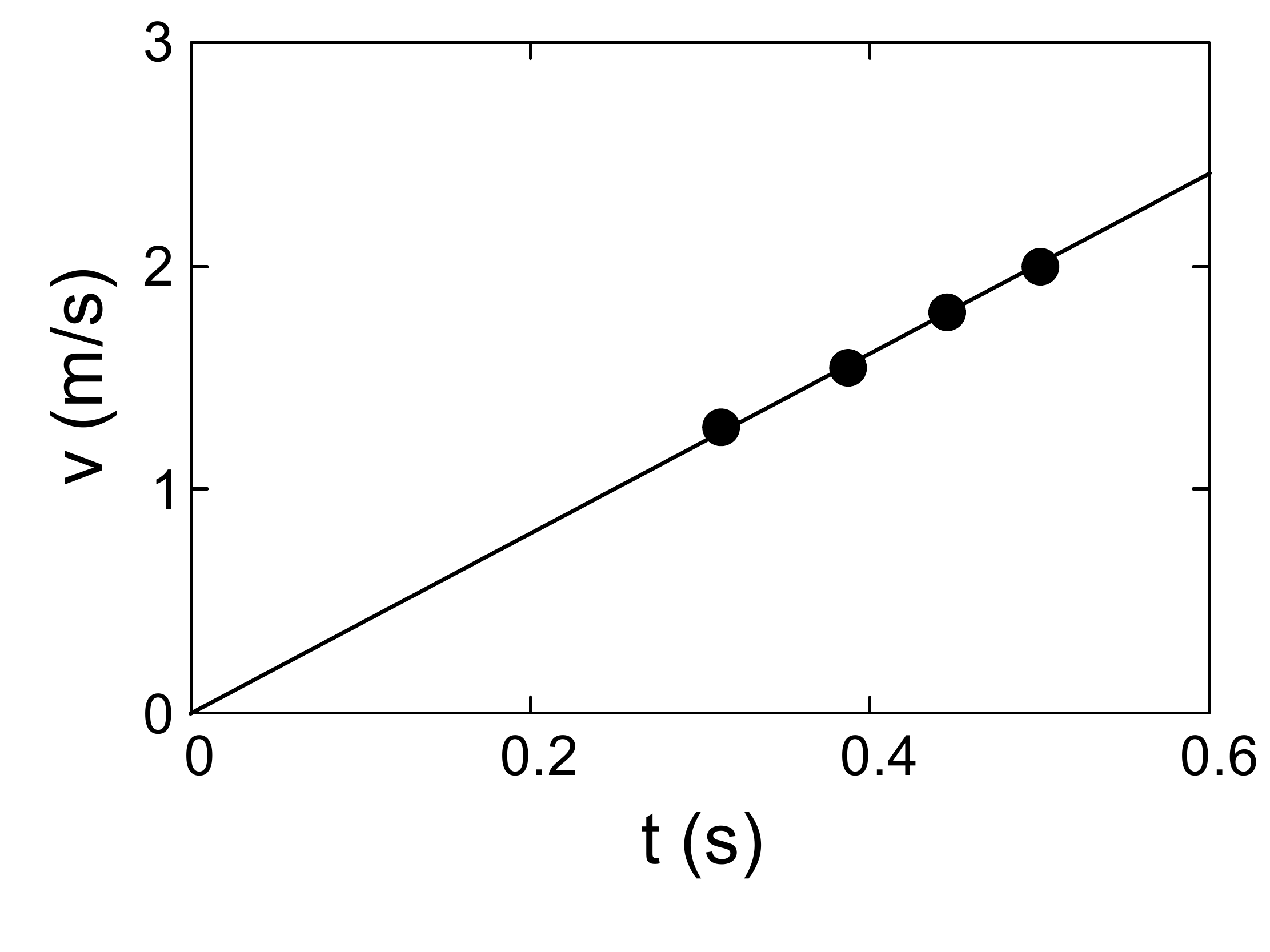}
\caption{Speed on the incline as function of time. The points are measurements with Galileo's device and the line represents a uniformly accelerated motion $v \propto t$ fitted to data.}
\label{vt}
\end{figure}

An example of measurements performed with Galileo's speedometer is in figure~\ref{vt}. The points show the instantaneous speed at the bottom of the incline for various descent times, corresponding to different starting positions. The data align quite well on a straight line $v \propto t$, indicating that the motion down the incline is uniformly accelerated.\footnote{It should be noted that the distances $x$ at which the data are taken were chosen by releasing the sphere from different points along the incline. Assuming, as did Galileo, that the sphere goes down the slope in the same manner irrespective of its initial position, we may interpret the different measurements as pertaining to a single motion starting at some point on the ramp.
}

\begin{figure}[htb]
\centering
\includegraphics[width=0.9\linewidth]{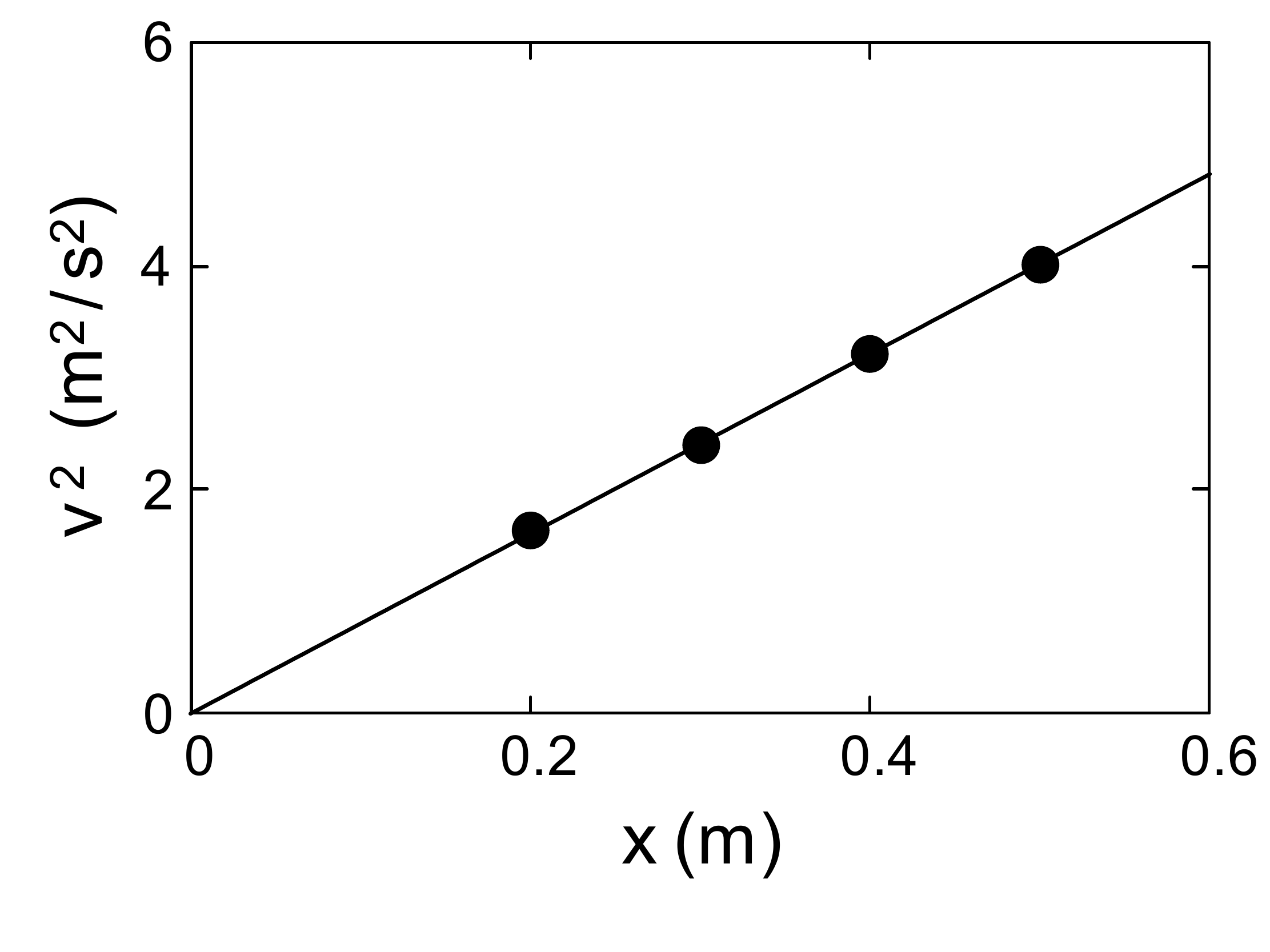}
\caption{Speed squared as function of distance traversed on the incline. The points correspond to the same measurements shown in figure~\ref{vt} and the line represents a fit of relation $v^2 \propto x$ to the data.}
\label{v2x}
\end{figure}

Data from the same measurements reveal how the speed at the bottom of the incline depends on the distance traversed on it. Figure~\ref{v2x} shows the speed squared as function of distance, and one notes that the data points follow closely the linear relation $v^2 \propto x$, as expected.

\begin{figure}[htb]
\centering
\includegraphics[width=0.9\linewidth]{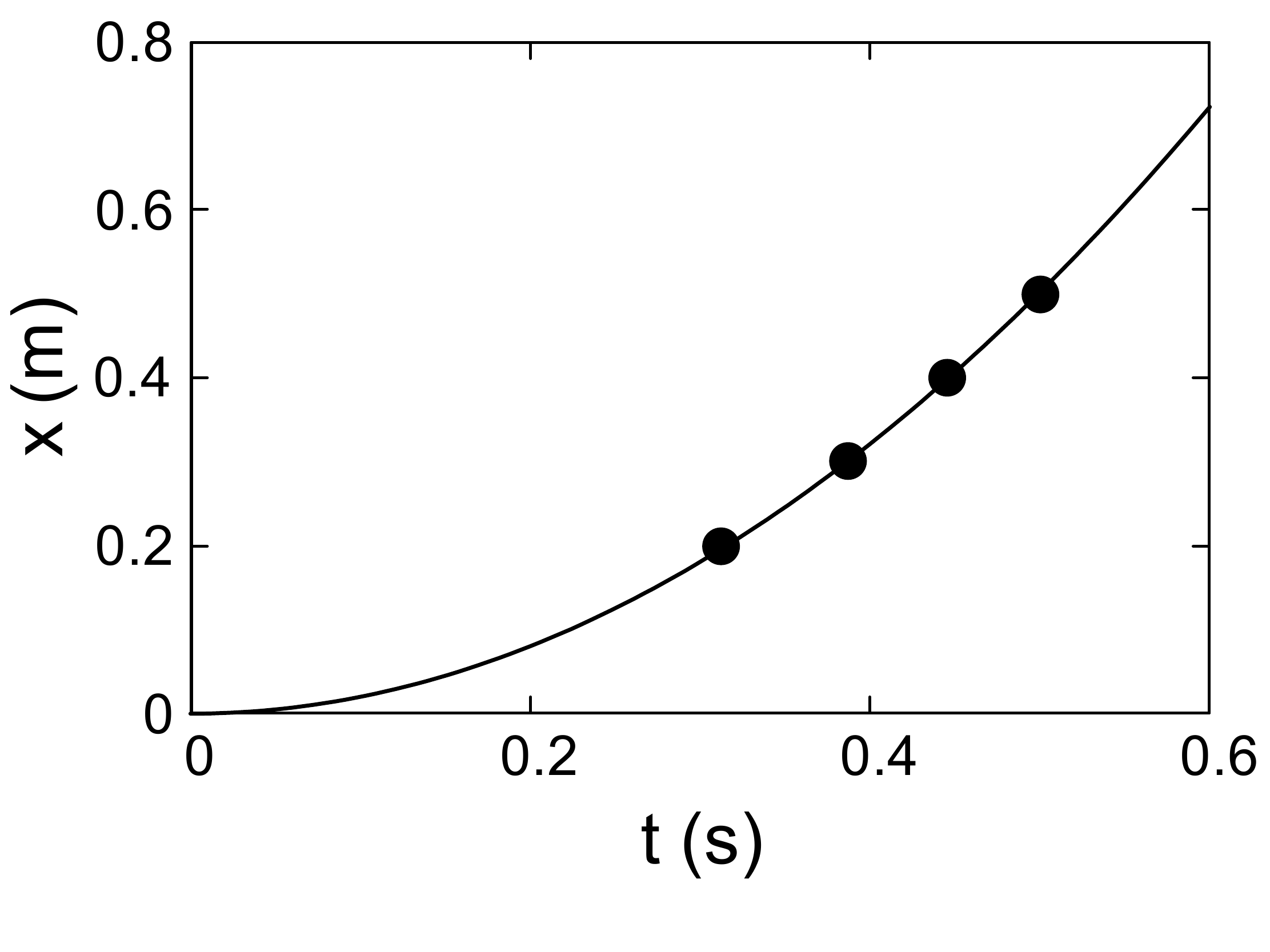}
\caption{Position on the incline as function of time. The measurements are the same as in figures~\ref{vt} and \ref{v2x}. The curve is a fit of $x \propto t^2$ to the data.}
\label{xt}
\end{figure}

The data we have collected also tell how the position on the incline depends on time. This is shown in figure~\ref{xt} and we see that the measured points are in good accord with Galileo's $x \propto t^2$  law, represented by the curved line.

Each of the representations used in the preceding figures provides a value for the acceleration $a$ on the incline. The results are consistent, as they should be: the $v=at$ fit in figure~\ref{vt} yields
$a = 4.02 \textrm{ m/s}^2$, 
using $v^2=2ax$ in figure~\ref{v2x} gives 
$a = 4.01 \textrm{ m/s}^2$,
and 
$x={\frac{1}{2}}at^2$ in figure~\ref{xt} leads to 
$a = 4.01 \textrm{ m/s}^2$. All the fitted values have uncertainty $\pm 0.02 \textrm{ m/s}^2$.

\begin{figure}[htb]
\centering
\includegraphics[width=0.9\linewidth]{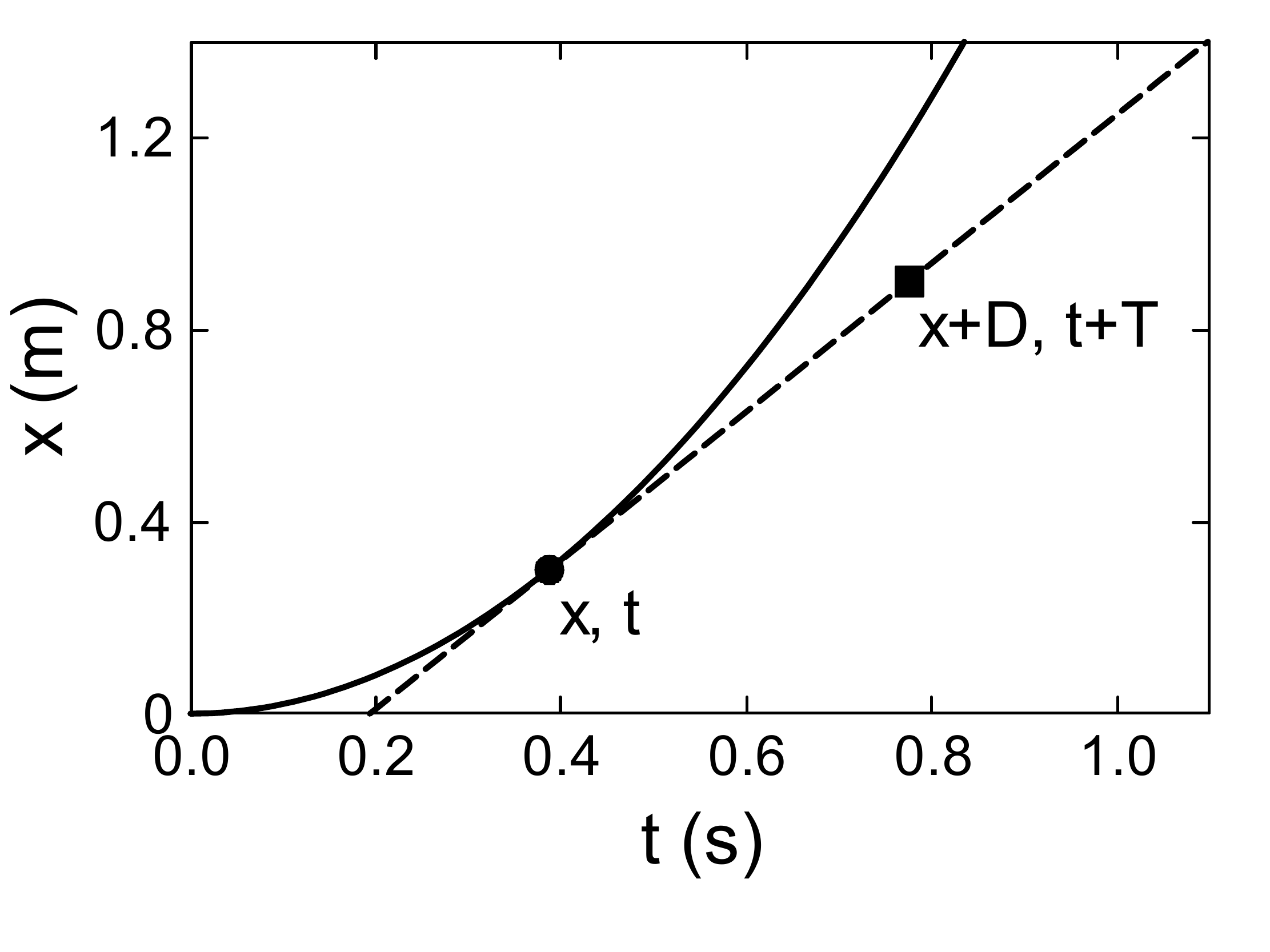}
\caption{Experimental realization of the Mertonian construct on a distance vs.~time plot. The uniform motion starting at $x,t$ defines a straight line that is tangent to the curve of non-uniform motion at this point, implying a connection between the Mertonian and calculus-based definitions of instantaneous speed.}
\label{tangent}
\end{figure}

An instructive connection between the Mertonian and calculus-based definitions of instantaneous speed can be established with the help of the distance vs.~time plot. 
To see this, the same curve that was fitted to the data of figure~\ref{xt} is plotted again in figure~\ref{tangent}, together with an arbitrarily chosen data point, of measured coordinates $x$ and $t$. Then we plot another point, with coordinates corresponding to the total distance traversed by the sphere on the tracks, $x+D$, and the total elapsed time $t+T$. The straight line joining these points describes the uniform motion idealized by the Mertonians (see figure~\ref{merton-v}) and made real in Galileo's speedometer. The interesting thing here is that, as far as one can tell visually,  the straight line is tangent to the curve at the chosen data point. Relating this tangent line to the derivative of position with respect to time allows for a fruitful discussion on the equivalence between the two definitions of instantaneous speed.

\section{Comments}

Physics education has taken much benefit from interactions with history of science. 
For instance, influential curricular innovations like the \emph{Harvard Project Physics} included historical content as a way of providing humanistic approaches to physical science~\cite{Holton}.
Also, physics education research revealed remarkable similarities between students' intuitive reasoning and theories of historical interest, like the pre-Newtonian notion of impetus in  mechanics~\cite{McCloskey}. 
In this paper we have seen how historical developments can help students overcome difficulties not unlike the ones faced by philosophers and scientists centuries ago. 
We suggest that the concept of instantaneous velocity, so challenging to students unfamiliar with calculus, can be made more accessible by ideas proposed by the Merton scholars and Galileo long before calculus was invented. The implementation of these ideas in experiments that can be carried out in a classroom makes it possible to ascribe an operational meaning to an otherwise abstract concept, and to obtain from measurements the laws of fall discovered by Galileo. 

It should be stressed that this approach is not meant to substitute the usual method of teaching instantaneous velocity, but rather to complement it.
We agree with A.~Arons~\cite[p~30]{Arons} that ``students must be given the chance to encounter the idea of instantaneous velocity slowly and with several episodes of cycling back to reencounter and reaffirm it as one proceeds through the study of kinematics and dynamics. Only a few students will absorb the concept on first encounter, but additional numbers break through in each subsequent episode.'' 
Perhaps history of physics can provide some of these episodes.

\bigskip\noindent Work partially supported by Capes, Brazil.

\end{document}